\documentclass[aps,prd,preprint,superscriptaddress,showpacs,showkeys]{revtex4}
\usepackage{amssymb,amsmath}
\usepackage{graphicx} 
\usepackage{dcolumn}  
\usepackage{epsfig}   
\usepackage{pstricks}
\usepackage{ifpdf}
\begin{document}


\title{Status of the wave function of Quantum Mechanics,\\ or,\\ What is Quantum Mechanics trying to tell us?}


%
%

\author{D-M. Cabaret}
\affiliation{Couvent Saint Etienne, Jerusalem}
\email[]{dominiquemarie.cabaret@gmail.com}
\author{T. Grandou}
\affiliation{Universit\'{e} de Nice-Sophia Antipolis,\\ Institut Non Lin\'{e}aire de Nice, UMR CNRS 7010; 1361 routes des Lucioles, 06560 Valbonne, France}
\email[]{Thierry.Grandou@inln.cnrs.fr}
\author{E. Perrier}
\affiliation{Couvent des Dominicains\\ 
1, Impasse Lacordaire\\ 
Toulouse\\ 
31400 France}
\email[]{perrier@revuethomiste.fr}


\date{\today}

\begin{abstract} 
The most debated status of the wave function of Quantum Mechanics is discussed in the light of the epistemological \emph{vs} ontological opposition. 
\end{abstract}


\maketitle

\section{\label{SEC:1}Introduction}

{\textit{`Now, what is the wave function indeed? Is it a concrete physical object?A kind of law of motion? An inner property of particles?A relation between space-time points? Or is it a summary of all of the information we have about the system under consideration?'}} \cite{Galchen}.
\par\medskip
{\textit{`I start from the consideration that the notion of observer is indispensable to the quantum theory, since it is the observer who identifies a quantum system and ascribes a state to that system (the description of all of the available information concerning the system)'}} \cite{Grinbaum}
\par\medskip
The examples above are just two among a lot of similar questionings concerning the famous wave function of quantum mechanics. This is because the fundamental point, that of the wave function status, either {\textit{epistemological}} or {\textit{ontological}}, is not settled yet. This long standing difficulty shows clearly enough that something really essential is still escaping our understanding of the quantum world, and that, in spite of spectacular improvements, the {\textit{quantum enigma}} persists.
\par\medskip
Examining this issue is the matter of the current article. In Section II, account will be given of the wave function epistemological dimension, an aspect barely questioned nowadays, when not taken for the only valuable one. An ontological dimension is not that easily refuted, though, as discussed in Section III. It is solely through an accurate enough grasp of the quantum reality that both dimensions, epistemological and ontological are not only reconciled but shown to be inseparable; this is the matter of Section IV. Intimately related to the wave function status is of course the issue of its reduction or collapse which is examined in Section V. In the light of the proposed understanding of the matter, some famous cases are revisited in Section VI, while elements of a conclusion are exposed in the seventh Section.

\section{\label{SEC:2} An indisputable epistemological dimension}

As stated in the Introduction, the epistemological status is taken for granted if not definitely indisputable: \textit{`First, I fully agree that a quantum state is no more than a state of knowledge, that there are no unknown quantum states ..'} \cite{private}. Several examples testify to this state of affairs. For the sake of illustration, the following ones can be proposed. Let us begin with considering a simple experiment of \textit{quantum optics} such as the one depicted in FIG.1.  It is assumed that a photon is emitted from a source at an instant $t_0$, and propagates towards a horizontal beam splitter by which, with equal probability, the photon will be either transmitted (symbolically represented by the state vector $|D_2\rangle$) or reflected (state vector $|D_1\rangle$). The photodetectors $D_1$ and $D_2$ are located at an equal distance to the beam splitter. After a lapse of time, $t\rangle t_0$, the wave function is written as the equally ponderated sum of the two possibilities the photon has to propagate through this experimental device,
\begin{equation}\label{3}
|\Psi\rangle=\frac{1}{{\sqrt{2}}}\big[|D_2\rangle+i|D_1\rangle\big]\,,
\end{equation}where in front of the state $|D_1\rangle$ appears that phase factor of $i$ a photon wave function picks up whenever reflected.
\begin{figure}[!htb]
   \includegraphics[width=\linewidth]{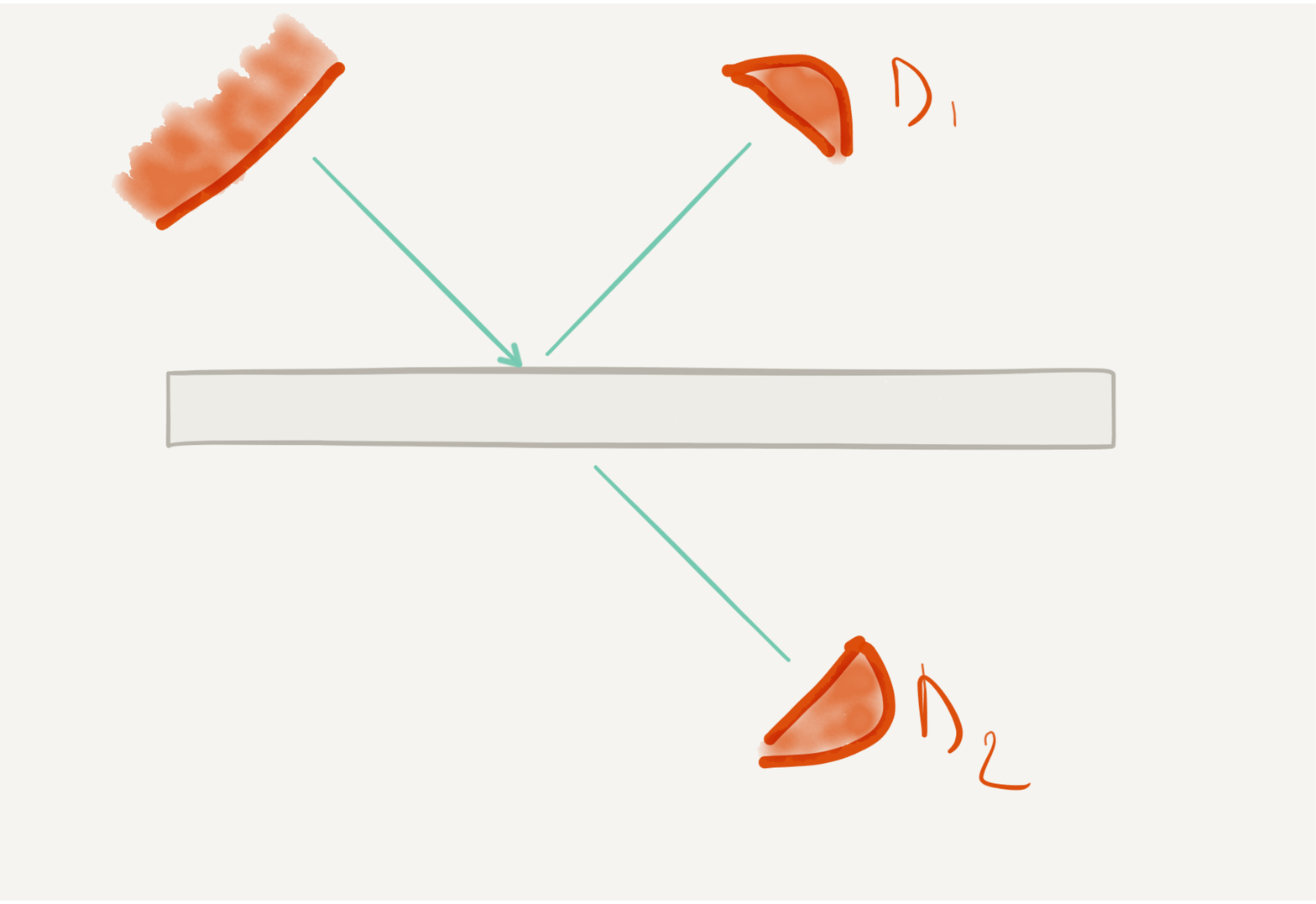}
   \caption{A source emits a photon reflected towards a detector D1, or transmitted to detector D2 by a beam-splitter}
   \label{fig1}
\end{figure}
The epistemological content of the wave function may be made even more cogent in the following two cases which, moreover, introduce the notion of \textit{wave function reduction}. Within the same experimental setup as the one of Fig.1, the vertical distance the photon has to travel before reaching $D_2$ is now substantially increased. As a result, if after a duration of $OD_1/c$ the photodetector $D_1$ has registered no `click', then one knows that the photon wave function reads now,
\begin{equation}\label{4}
|\Psi\rangle=|D_2\rangle.
\end{equation}
That is, a \emph{wave function reduction mechanism} has operated even though no measure has ever been performed on the system. 
\par\medskip
A second example of this kind is furnished by an experimental device where a radio active atom is surrounded by a spherical detector $D_1$, but not over the whole $4\pi$ \textit{steradians}: A hole, a solid angle $\Omega$ wide, allows a detector $D_2$ to be located outside of the enclosure, as depicted in FIG.2, \cite{Laloe}.  It is assumed that the radio active source emits in an isotropic way. In a certain time range, the spherically symmetric wave function thus reads (using intuitive enough notations),
\begin{equation}\label{5}
|\Psi\rangle={\sqrt{\frac{4\pi-\Omega}{4\pi}}}\,|4\pi-\Omega\rangle+{\sqrt{\frac{\Omega}{4\pi}}}\,|\Omega\rangle\,.
\end{equation}After a long enough time, in case of no detection by $D_1$, the initial wave function (\ref{5}) gets automatically reduced to its sole $|\Omega\rangle$ component with probability $100\%$, and wave function reduction has therefore taken place, while no measure or any sort of perturbation has ever impacted the system. 
\par
\begin{figure}[!htb]
   \includegraphics[width=\linewidth]{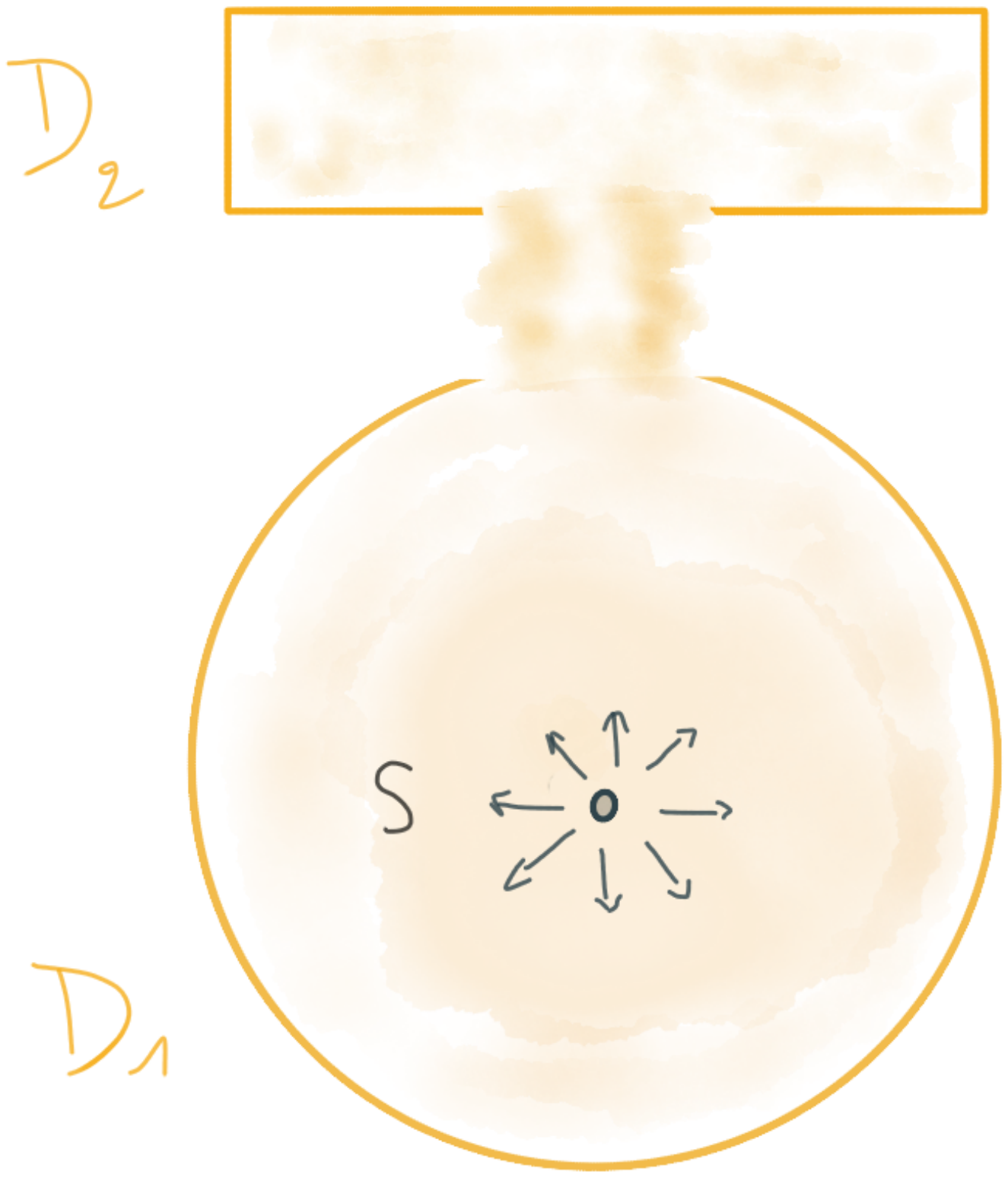}
   \caption{Isotropic emission of a photon by a radio-active source comprised within the enclosure $D_1$}
   \label{2loops}
\end{figure}
As recalled in  \cite{Laloe}, it is important to discard the analogous situation where the source would have emitted in the direction of $D_2$. In this latter instance, the no detection event at $D_1$ only discloses this initial direction of emission, and nothing else is to be learned out of this experiment. Now, so long as \textit{hidden variables} explanations are excluded (see Section V), the quantum mechanical situation is very different. This is because no direction ever pre-exists to its detection, which itself is the matter of an interaction process. This emphasizes the peculiarity of the current experiment where a bunch of directions is revealed by a \emph{no detection event}, that is in the absence of any detection device. As noticed in \cite{Laloe} where this experiment is signalled a `rather curious' one, this result conflicts the standard \textit{Copenhagen interpretation} according to which wave function reduction results of the \textit{quantum jumps} induced by the interaction inherent to any detection or measure.
\par\smallskip
Now, it is important to remark that if the wave function is to be restricted to an epistemological content only, the passage of (\ref{5}) to $|\Omega\rangle$, or of (\ref{3}) to (\ref{4}), is in no way problematic. Rather, it is to be thought of in terms of an updating of our knowledge of the quantum system state, quite in line again with the opening quotations of Section \ref{SEC:1} or the recent \textit{Bayesian interpretation} of Quantum Mechanics \cite{Bayes}. However, a much simpler interpretation of this `curious experiment' will be given in Section VI.

\par\medskip 

A more involved and interesting situation is that of a \textit{Mach-Zehnder Interferometer} in which a photon is entering, as represented in FIG.3, \cite{Gerry, Grangier}. After having passed the beam splitter $BS_1$. the photonic wave function will read as,
\begin{equation}\label{1}
|\Psi\rangle=|\mathrm{After}\, BS_1\rangle=\frac{1}{{\sqrt{2}}}\big[|CW\rangle+i|CCW\rangle\big]\,,
\end{equation}with the symbolic notations ($CW,\,CCW$) referring to clockwise and anti-clockwise photon's paths in the \textit{Mach-Zehnder Interferometer}. Later on, the transmitted and reflected parts of the initial beam hit upon another beam splitter, $BS_2$. Therefore, after $BS_2$, the state vector $|CW\rangle$ goes into the new state vector $(1/{\sqrt{2}})\left( |D_1\rangle+i|D_2\rangle\right)$ and the state vector $|CCW\rangle$ into $(1/{\sqrt{2}})\left( |D_2\rangle+i|D_1\rangle\right)$. In the end, the full wave function after both $BS_1$ and $BS_2$ reads,
\begin{equation}\label{2}
|\Psi\rangle=|\mathrm{After}\, BS_1\,\mathrm{and}\, BS_2\rangle=\frac{1}{{{2}}}\big[-|D_1\rangle+i|D_2\rangle\big]\,,
\end{equation}

\begin{figure}[!htb]
   \includegraphics[width=0.7\linewidth,angle=90]{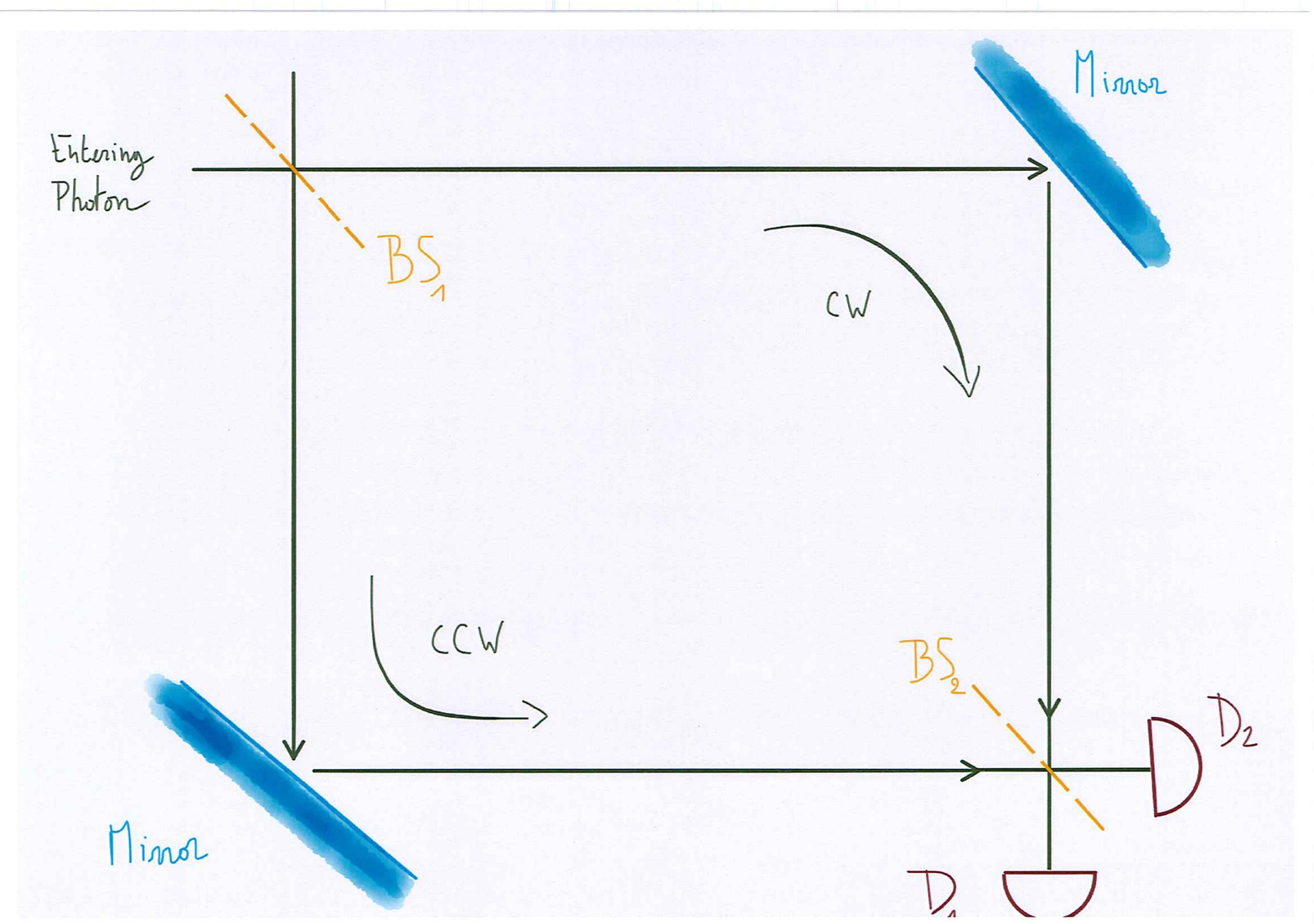}
   \caption{A source emits a photon eventually detected by D1 or D2.}
   \label{figg1}
\end{figure}

Clearly, at each of these steps, forms like (\ref{3}), (\ref{1}) or (\ref{2}) just account for the whole set of possibilities a photon has to propagate through the Mach-Zehnder interferometer pictured in FIG.3. Experiments of this kind allow one to write down the complete \emph{catalog} of the possibilities offered to the propagating photon, each of these ponderated by a complex number which is the \textit{amplitude of probability} of the corresponding possibility. In here, one recovers the basics of the second quotation opening the Introduction.

 \section{\label{SEC:3} An ontological dimension}
 
 {\textit{`The wavefunction is the complex distribution used to completely describe a quantum system, and is central to quantum theory. But despite its fundamental role, it is typically introduced as an abstract element of the theory with no explicit definition. Rather, physicists come to a working understanding of the wave function through its use to calculate measurement outcome probabilities by way of the Born rule  \cite{Lundeen}.'}}
 \par\medskip
 This new quotation is interesting also and displays an evolution, powered by `working understanding', towards a pragmatic attitude in front of the quantum enigma. Such a philosophical stand point receives a recent and remarkable expression in \cite{Bachtold}, for example. Now, of course, it does not elude the ontological issue whose questioning can be phrased as follows.
 \par\medskip
 \emph{`How could it be, in effect, that so efficient a mathematical tool as is the wave function, enabling one to completely describe a quantum system, had no deeper relations to the very {\textit{nature}} of the quantum system, making of it some form of physical property of that system?'} At the level of pure principles, one cannot express in a better way the ontological questioning, challenging a wave function interpretation that would be exclusively epistemological \cite{Wolfgang}.
 \par\medskip
 For physicists, there is no doubt that an ontological content for the wave function can be agreed on if in some way or other, it is possible to measure it. If, given a single quantum system it is possible to measure its wavefunction, then this wavefunction is a physical reality of the corresponding quantum system, and not just the knowledge we have about it. 
 \par\medskip
 More than twenty years ago, it is along these lines of thinking that one attempted to establish the measurable character of the state vector on a simple enough quantum system~\cite{Aharonov}. The proof however couldn't be taken as convincing enough because it relied on the assumption that measures could be performed on the wavefunction while preserving it the same as it was before any measurement be performed \cite{Unruh}.
 \par\medskip
 Now, substantial improvements have been realized since then. 
 \par
 - So-called tomographic methods have been introduced {\textit{ `(..) which estimate the wave function most consistent with a diverse collection of measurements.'}}, \cite{Lundeen}. But as noticed in the same analysis, such indirect methods aggravate the problem of defining the wavefunction. Another method has therefore been devised, which is direct in the sense that it doesn't involve the sets of measurements and computations inherent to tomographic approaches.
 \par
 - The method consists in a sequential measurement of two complementary variables of the quantum system, the first of them realized by way of {\textit{weak measurement technics}} \cite{Lundeen}. An interesting feature of a weak measurement is that, in some limit, it doesn't disturb a subsequent normal (or strong) measurement of another observable. The example of a single particle can be given. Considering the weak measurement of position, associated to the hermitian operator $A\equiv \Pi_x\equiv |x\rangle\langle x|$, followed by a strong measurement of momentum giving value $p$ to the momentum operator $P=i\hbar\, \partial/\partial x$, one gets,
 \begin{equation}\label{weak}
 \langle\Pi_x\rangle_W=\frac{\langle p|x\rangle\,\langle x|\Psi\rangle}{\langle p|\Psi\rangle}=\frac{e^{ipx}\,\Psi(x)}{\Phi(p)}\end{equation}where $\Phi(p)$ is the Fourier transform of $\Psi(x)$, and the subscript $W$ is to remind a weak measurement. In the limit of $p=0$, Equation(\ref{weak}) gives therefore $\langle\Pi_x\rangle_W=k\Psi(x)$ with $k=1/\Phi(0)$. The weak measurement of the position, followed by a normal measurement of the momentum has produced a result proportional to the wavefunction of the particle at $x$. It therefore remains to scan the weak measurement through $x$ in order to `obtain' the complete wavefunction. Along this procedure the real and imaginary parts of the state vector $|\Psi\!\!\rangle$ are directly readable out of the measurement apparatus, and an unambiguous operational measurement of the quantum state is achieved, making it clear that the wave function is a real physical object.
 \par
 - Inspired by remarkable experimental achievements \cite{Raimond, Guerlin}, bearing on quantum trapped fields, indirect {\textit{quantum non-demolition}} measurements can also be used so as to reconstruct the initial photon distribution of a resonant electromagnetic cavity in a successful way \cite{Bauer}. Quantum non-demolition measurements are those which collapse the state vector in some eigenstate of the free evolution of the system considered, while indirect measurements are performed on {\textit{probes}}, which are entangled to the system considered by way of a short time interaction with the latter.
 \par\bigskip
 To sum up, remarkable technical improvements have been realized, combining highly preserving measurement procedures with entanglement to probe-systems on which measures can be performed, while leaving untouched the system of interest. In the same time, it is worth pointing out that all of these technics can be fully transcribed within the quantum formalism itself. In the end, they lead to an operationally well defined wave function, as well as to its possible measure on a given quantum system.
 \par\medskip
In the end, if the wave function is in any way a measurable entity, as it turns out to be, then it must be concluded that it is definitely a \textit{real} property of a quantum system, and in this way, the ontological status of the wave function is established.

 \section{\label{SEC:4} Synthesis}
 
 \subsection{\label{A} Introducing a necessity}
 
 As we have just seen, arguments in favor of an epistemological status of the wave function are sound and indisputable. Now, so seem to be also the arguments establishing the physical reality of the wave function itself, that is, its ontological status. Reconciling both aspects looks like an impossible task, as evidenced by the scientific literature amply.
 \par
The {\textit{quantum enigma}} which this dilemma examplifies, has to do with a circumstance which has no equivalent in the whole history of classical physics. In order to grasp what is all this about it may be helpful to recall an evidence, easily forgotten, though crucial \cite{Daujat},
\par
{\textit{Theoretical physics is a mathematical analysis of physical reality by means of numbers}} 
\par\noindent
From this point of view, of course, quantum theory doesn't depart from classical physics. But it helps understanding in retrospect that classical physics is {\textit{phenomenological}} \cite{mlb2}. That is, to a classical system, whose full complexity is now revealed through its quantum elementary components, physics substitutes a convenient idealisation which sorts out a few physical measures out of a real complex reality,
\par
{\textit{The seeming completeness of classical physics betokens the fact that we are dealing not so much with physical entitys as with convenient abstractions \cite{Wolfgang}}}.
\par
 Thermodynamics, with its measures of temperature, pression, entropy and other state functions, whose character of mean values is evidenced by {\textit{statistical mechanics}}, is such an example of a phenomenological theory. And for a wide range of physical concerns, it is unnecessary to recourse to a deeper level of description that would be provided by the non-zero temperature realisations of quantum field algebras \cite{mlb4}. 
\par\medskip
And thus, theoretical knowledge deals not so much with the whole thing, as it is, than with its simplified apprehension. In a nutshell, all non relevant aspects can be bracketed out during the classical analysis by numbers. Classical physics has traded the nature of physical systems for very simplified idealisations with the indisputable  advantage that `we know what we talk about' : Not the systems in themselves, but the systems which, in certain conditions and for certain usages do not drift too much apart from their idealisations, properly formalized. It is worth recalling here that this attitude belongs to the birth certificate of modern Science such as announced by R. Descartes in 1644 \cite{Descartes}. 
\par\medskip
Such is not the case of the quantum world. The quantum theory is able to describe the quantum world very well, individual systems included, and is never taken at fault. Now, at variance to the classical situation, the core of the quantum enigma is this time that in contradistinction to classical physics, `we do not know what we talk about'.
\par
This is because, as evidenced at length, quantum objects {\textit{are}} not in the same way as classical objects {\textit{are}} \cite{Grandou}. This, once, was put forth by R.P. Feynman with some humour, {\textit{`All quantum objects are crazy, but they all behave in the same crazy way'}}. The point is that one needs to know how quantum entities {\textit{are}}, taking seriously what the quantum world tells us about itself.

  \subsection{\label{B} A re-orientation of the opposition epistemological/ontological}
How could there be any knowledge of an entity that would be a knowledge solely? It must be realized that such a position, which is that of a radical dissociation, amounts to posit that there is eventually no access to {\textit{reality}}, an interpretation that can be often found in the literature \cite{LaRecherche2014}; even though in this extreme case, as in the case of the famous {\textit{Interpretation of Copenhagen}}, one should rather speak of a {\textit{non-interpretation}}, as A. Legget suggested \cite{Legget}. Science would build up algorithms, some of them luckily endowed with enough efficiency, but deprived of any sound relation to reality.  Probably under the influence of a series of theorems concerning the self-limitations of logical formal systems, even a philosopher like J. Maritain suggested that \textit{`.. theoretical physics is but an art of exploration of reality by means of checked myths'} \cite{Maritain}. 
\par\medskip
On the contrary, as will be argued shortly, the wave function tells us something of the reality of the quantum system it describes so efficiently, and is not solely the knowledge one has about it. This, however, will not be in the sense of Section \ref{SEC:3}.
\par\medskip
Now, are there realities which can help unravelling a bit how quantum objects \textit{are}? That is realities which we understand through the double, epistemological and ontological modes. Let us bring about the case of a seed. In a seed, a whole panel of grown up trees are contained in an inchoative manner. Even though, a given seed will ever produce a unique tree out of the whole set of possible trees. What is to be stressed here, is the opposition between a determined {\textit{actual}} state, and the fact that this actual/realised state was present already in the whole set of possibilities the seed entailed. This case exhibits a mode of reality which is plainly real though not actual, being a whole set of possible actualisations one of which only will get realised.
\par
As the seed itself, this \textit{potentiality} must be regarded as real. In particular, it departs from a pure imagination one could form concerning the possibilities the seed comprises. In the seed in effect, an infinity of other possibilities are excluded. The range of possibilities in the seed, is a constrained range, real, even though a mere potentiality with respect to the full possible actualisations. This case exhibits two states of one and the same entity, its potential state and its actual or realised state.
\par\medskip
When it comes to the wave function, the point of view of Section \ref{SEC:2} is correct, as it points out the indisputable descriptive content of the wave function considered within the masterful setting allowing one to reach an explicit writing of this wave function; what can be grouped under the generic term of a \textit{preparation} \cite{Peres} (See the second opening quotation of Section \ref{SEC:1}). A preparation in effect, constrains the set of possible results without determining any of them in particular. This accounts for the so-called epistemological dimension, or the \emph{catalog aspect} of the wave function. 
\par\medskip
However, sticking to this interpretation in an exclusive manner, as it is widely agreed on nowadays, another aspect is totally missed, which is crucial. One simply overlooks that a catalog only requires \textit{commas}, \textit{semicolons} and \textit{points}, and that a wave function is never written in this way. A wave function is not simply a catalogue of possibilities. It is a linear combination of possibilities whose complex valued coefficients have  squared absolute values adding up to unity: Be it at a formal level only, this cannot be mistaken for a catalog.
\par\medskip
Writing a (mathematical) statement such as (\ref{3}),
\begin{equation*}\label{}
|\Psi\rangle=\frac{1}{{\sqrt{2}}}\big[|D_2\rangle+i|D_1\rangle\big]\,,
\end{equation*}and postulating its unitary evolution, 
cannot be exhausted by an interpretation in terms of catalogs. It is nothing less than a \textit{judgement of being}, since at the very least, (\ref{3}) is the affirmation that a \textit{unity} exists and persists under both unitary evolution and collapse, supported by an entity whose behaviour is such that it is adequately described by this linear combination of possibilities. Moreover, taken as actualised, these possibilities are mutually exclusive (the discussion is here restricted to the finite dimensional case), which indicates that the linear combination bears on potentialities, not on realisations.
\par
 In positing such a statement as a wave function, it is obvious that physics takes the risk of being denied by experimental facts. As long known, on the contrary, far from being denied, the relevance of `wave function statements' to the experimental behaviours of quantum systems is firmly established.

The interest of analyses distinguishing between the reality of actual states and the reality of potential ones, is that they are able to account for the relevance of the wave function \emph{as it is}, its very structure testifying to `the level of reality of what we talk about'. In contrast, a purely epistemological wave function would not be able to account for the operatorial efficiency of the wave function description in all possible cases \cite{Zeh}.
 \par
 One may observe that the report (rather than `the measure', more problematic an issue to be addressed elsewhere) of a given realised actual state offers a \textit{datum}, an experimental fact quite similar to the measured data of classical physics: One fixed and stabilized \cite{quirky} observation, a real number whose probability of occurrence is given by the Born's rule which, at the formal level which is its own, explicitly projects the potentiality of the wave function on this particular actualisation. 
Along this line of thinking the quantum realm, a few more key-points will reveal helpful.
\par\medskip
$\bullet$ In a linear superposition or in an actual state, there is but one and the same subject. Unitarity is the conservation of the overall probabilities of all possible realisations. Once a given realisation comes about, a formal discontinuity takes place as the series of complex-valued probability amplitudes gets reduced to a unique real-valued coefficient of $100\%$. This discontinuity however does not compromise the continuity/conservation of the underlying subject, since on the contrary, it is in this way that the latter is asserted through the passage of potential to actual.
\par
$\bullet$ The mutual exclusivity of act and potency. Note that potentialities can only be apprehended through actualities, and that only the latter can be measured, in the classical case, or induced by the measurement process in the quantum case. The mutual exclusivity of act and potency means that measures can be conceived and performed only by getting out of a potential state.
\par
$\bullet$ Act is prior to potency. This is true of knowledge. As just quoted above, potentialities are meant through acts and it is impossible to apprehend a pure potentiality in itself. It is true also at an ontological level. As a matter of fact, what can be thought of as being potentially, is always thought of an object possessing a certain degree of actuality. All of the properties that are said potential are properties of an object which is in a certain actual state, like the seed in the analogy used above. This is a crucial point. As will be dealt with in details elsewhere, this principle extends down to the so-called \textit{materia prima}, the existing reality standing for the lowest level of all possible actualities \cite{PEs}.
\par
$\bullet$ A reality in a given state of potentiality does not leave this state of potentiality on its own. An action is needed, operating upon it, so as to induce such a transition.

\section{\label{SEC:4a} On the wave function reduction mechanism}

  \subsection{\label{Aa} The issue}

If there is an issue related to the wave function status it is no doubt that of its most debated \textit{collapse} or \textit{reduction}. In textbooks, this reduction often takes the form of a supplementary axiom of Quantum Mechanics ($QM$). However, the wave function collapse has a somewhat \textit{ad hoc} aspect as it seems to challenge the unitary evolution prescribed by the Schrodinger equation. For these reasons the wave function collapse has always motivated a lot of attempts at either eliminating it from the theory (Everett's many worlds interpretation is such an attempt where only the unitary reversible evolution controlled by the Schrodinger equation is retained), or deducing it from more axiomatic or intuitive bases \cite{quirky}. In \cite{Zurek}, for instance, the \textit{no-cloning theorem} is used to prove that if a measurement was to leave a quantum system on which it is performed in any other state than an eigenstate of that measurement, then no stable measure could ever be envisaged: Immediately after the first measurement, the same measurement performed again could deliver another result.
\par
At face value, though, the above argument looks rather like an {argument of internal coherence}: Given the basic axioms of $QM$ on the basis of which the no-cloning theorem applies, the argument displays the adequacy/compatibility of the wave function reduction axiom rather than its logical reduction to previous axioms. As an evil rarely comes about alone, the wave function reduction axiom comes along with the famous \textit{Born's rule}, giving, out of a series of identical measurements, the probability distributions of all possible realisations. 
\par
So far, it is worth noticing that attempts at deriving the Born probabilities out of an Everett's many worlds interpretation have failed, and this can be taken as an indication that a real collapse of the wave function takes place in deeds \cite{Giacosa}.
\par
Along this line of thinking though, several mechanisms have been proposed over the past thirty years to account for a physical explanation of the wave function reduction. Most of them rely on the assumption that the \textit{linear} Schrodinger equation is but an approximation to a more complete equation involving \textit{some} non-linear extensions \cite{GRW}. These models now, real physical space-time mechanisms, have to become more and more involved, include gravitation, whose unification to quantum physics is not available, and have ultimately to include also the observer's minds in order to.. fail again and again. A most interesting account of all these attempts can be found in \cite{Laloe}.

  \subsection{\label{Ba} Real time observation of the collapse}
 The excellent presentation given in \cite{Laloe} is followed here. In the early times of \textit{QM}, and even since them, for a large number of experiments involving \textit{QM}, measured quantities correspond to an averaged sum over a very large number of elementary systems (\textit{e.g.}, particles), of one and the same microscopic observable. This, by the way, is the state of affairs which long inspired the agreed on interpretation of $QM$,
 \par
 \textit{`.. the wave function describes in no way the state of a sole system; it has to do with a large number of similar systems in the sense of statistical mechanics .. if the wave function furnishes statistical data on measured observables, the reason is that the wave function does not describe the state of the unique system!'} (A. Einstein, \cite{Einstein}).
 \par
 This interpretation of $QM$ has persisted long after 1936, until the relevance of the wave function description to elementary systems became unquestionable.  In the case of Einstein's 1936 point of view, the mean values that are relevant to these situations are satisfactorily described by the continuous Schrodinger equation and there is no need to recourse to a wave function collapse axiom and/or Born's rule. As advocated in \cite{Giacosa} however, it is interesting to note that this could also apply to most elementary quantum optic systems: Not really necessary, the wave function reduction postulate would just appear as an effective way of description. But a closer look taken at such elementary cases reveals the same difficulty as mentioned above:
 How, then, to derive the Born rule probability distributions of random quantum events \cite{Giacosa}? Or to put it more bluntly, Schr\"odinger equation alone doesn't seem to suffice. Hard to circumvent in any way, one encounters here a point of resistance which could very well be the indication of a physical reality one has to take seriously and strive to understand. 
 \par\medskip
 The physical reality under consideration comprises the case of \textit{quantum jumps}. Through experiments carried out over elementary and individual quantum systems, distinct groups have been able to attest to the reality of these quantum jumps, and this under `Quantum Non-Demolition Measurement' circumstances, what is even more illustrative \cite{Laloe2}. In all of these examples, quantum jumps show up random and instantaneous and this is why it is worth considering things from the most achieved and available point of view, that of the covariant quantum theory of fields.
 
   \subsection{\label{C} Relativistic considerations}
   The fact that quantum jumps are instantaneous is far from insignificant. The point however is that \textit{instantaneity} is not a relativistic invariant notion. Most treatises on the matter are carried out within ordinary \textit{QM} because, as ordinarily motivated, \textit{ it is sufficient!.}. At the level of principles now, examining physical realities at such a deep \textit{resolution scale}, it may be somewhat bold to ignore that our Universe is not shaped by $\hbar$ solely, but by $\hbar$ and $c$ inseparably. This is why it may be appropriate to examine the collapse in the context of Quantum Field Theories that have proven to offer a most efficient descriptions of the quantum world \cite{Grandou} for a whole fan of its physical manifestations \cite{PEs}.
   \par
   Creating a particle at space-time point $x'=(x'_0, {\vec{x'}})$, a \textit{Klein-Gordon wave function} spread over the whole forward cone of summit $x'$. Annihilating the particle at some later point $x=(x_0,{\vec{x}})$, then the wave function disappears from spacetime. Within the $QFT$ formalism this can be displayed as follows \cite{Kleinert}. The Klein-Gordon field reads,
   \begin{equation}\label{KG}
   G_F(x-x')=\langle T\varphi(x)\varphi^\dagger(x')\rangle=\sum_p\frac{1}{2V\omega_p}\,e^{-ip\cdot(x-x')},
       \end{equation}where the scalar quantized field $\varphi$ lives in a spacetime taken as a big box of volume $V$. The particle density at a spacetime point $x''$, lying slightly later than the later time spacetime point $x$, can be accessed to by inserting the current density operator $\rho\sim \varphi^\dagger(x'')\varphi(x'')$ in the left hand side of (\ref{KG}); that is, by considering the \emph{Green's function},
   \begin{equation}\label{KGcollapse}
   \langle T\rho(x'')\varphi(x)\varphi^\dagger(x')\rangle=\langle T\varphi^\dagger(x'')\varphi(x'')\varphi(x)\varphi^\dagger(x')\rangle\,,
   \end{equation}where $T$ is the standard time-ordering prescription for products of field operators (somewhat redundant here). In view of the time ordering, a \emph{Wick Theorem} expansion of (\ref{KGcollapse}) yields as an obvious result,
   \begin{equation}\label{wick}
   \Theta(x''_0-x'_0)\,G_F(x''-x')\times  \Theta(x_0-x''_0)\,G_F(x''-x)=0\,,\end{equation}where $\Theta(x)$ stands for the \emph{Heaviside step function}, $\Theta(x)=0$ for $x< 0$, and $\Theta(x)=1$ for $x> 0$ (and to be complete, $\theta(0)=1/2$). That is, the Klein Gordon field produced at spacetime point $x'$ has completely disappeared from spacetime at any time $x''_0$ arbitrarily close to $x_0$, provided it comes later, $x''_0> x_0$. Since time ordering is preserved by the Lorentz ortochronous group, one could think that the instantaneous collapse of the wave function is established. And it is. 
   \par
   Now it is established within a framework, $QFT$, which does not and can not describe such a mechanism as the \textit{creation} or \textit{annihilation} of a particle. An equation such as $a^\dagger(\vec{k})|0\rangle=|\vec{k}\rangle$ doesn't describe a process, the creation of a quantum endowed with a momentum $\vec{k}$. It just accounts for it in the manner of \textit{a change of state}, a perfectly controlled notion of theoretical physics contrarily to the notions of creation and annihilation which totally escape its domain . It is important to keep this distinction in mind not to be fooled by words, and we owe the explicit formulation of this point to B. d'Espagnat \cite{dEspagnat, Grandou}. 
   \par
   In a sense, in a $QFT$ context, the immediate collapse of the wave function is simply built-in and the proof given above is accordingly \emph{circular}. However, the point is that $QFT$ expectations are verified by experimental data, and this to an impressive rate of accuracy. So, if not really demonstrated in full rigour, the instantaneous collapse of the wave function is strongly favoured by the covariant $QFT$ analyses.
   \par\medskip
   Before proceeding it is worth mentioning the case proposed in \cite{Kleinert}, right after equations (\ref{KG})-(\ref{wick}), which are basically its own. A two-particle state is considered $|\varphi(x'_1)\varphi(x'_2)\rangle$, and one focus on the Green's function,
   \begin{equation}\label{2to1}
   G_F(x'', x_1, x'_1, x_2, x'_2)=\langle T\rho(x'')\varphi(x_1)\varphi(x_2)\varphi^\dagger(x'_1)\varphi^\dagger(x'_2)\rangle\,,\end{equation}where again, the one-particle current operator $\rho\sim \varphi^\dagger(x'')\varphi(x'')$ is here inserted to keep track of a possible one-particle behaviour: That is, depending on the relative chronological order of $x''$ with respect to $x_1$ and $x_2$, (\ref{2to1}) allows one to learn about \emph{the one-particle content of the two-particle wave function collapse} \cite{Kraus}, a statement which may sound a bit weird \emph{a priori} while being highly significant indeed.
   \par
    To stick to our concern and to the extended one proposed in \cite{PEs}, this is a place where to point out a pre-eminence, ontological, of \textit{the field physical reality} as compared to that of the associated particles: The two-particle wave function just alluded to above is but a particular state of the field attached to that particle, prior to any of the one or two-particle contents which may result from its various collapses/measures. 
    \par
    A quite recent and astounding discovery, going beyond the earlier case of \cite{Kraus}, can make it easier to catch \cite{Georgi}.  The simplest case of two massless particles is used to illustrate the point, though holding true for any other number $n\geq 2$. In some cross-section calculations (used to predict some final experimental outputs) one has to integrate over the allowed phase space. Let us suppose that one is in a situation where the two final particles are not observed, so that only their total energy-momentum $P=p_1+p_2$ is known. One can combine the \textit{phase spaces} measures for the two massless particles, 
     \begin{equation}\label{p1p2}
d\rho(P)\equiv\prod_{i=1}^{2}d\rho(p_i)=\prod_{i=1}^{2}\delta(p_i^2)\,\Theta(p^0_i)\,\frac{d^4p_i}{(2\pi)^3}\,,\ \ \ \ i=1,2\,,    
    \end{equation}into the phase space measure relevant to their combination, $d\rho(P)$. This reads,
       \begin{equation}\label{P}
d\rho(P)=\int \prod_{i=1}^{2}\delta(p_i^2)\,\Theta(p^0_i)\,\frac{d^4p_i}{(2\pi)^3} \ \delta^4(P-\sum_{i=1}^2 p_i)\,{\mathrm{d}}^4P \,.
    \end{equation}With respect to the right hand side of (\ref{p1p2}) absolutely no change has been brought about because of the identity,
    \begin{equation}
   \int{\mathrm{d}}^4P\,\, \delta^4(P-\sum_{i=1}^2 p_i)=1\,,\end{equation}which means that the right hand side of (\ref{p1p2}) has simply been multiplied by $1$. The most remarkable fact is now that (\ref{P}) can be re-written to give,
    \begin{equation}\label{remarkable}
d\rho(P)=\frac{1}{8\pi}\Theta(P^2)\Theta(P^0)\frac{d^4P}{(2\pi)^4} \,.
    \end{equation}
In the right hand side of (\ref{remarkable}) nothing ever more refers to the constituent particles, and due to the Heaviside distribution of $\Theta(P^2)$ the system can have any mass squared, $P^2$, and cannot itself be thought of as a particle. Even though the two-particle origin of the resulting system is known from the onset, (\ref{remarkable}) means that the two particles as such are not necessarily observable in all cases. The right hand side therefore refers to a deeper \textit{substratum} of the physical reality, the one which in some cases is able to  sustain the two particle realisations. As advertised above, this deeper level of the physical reality is that of the field attached to the particles, and it is really remarkable that the formalism is itself able to testify to this structure of the physical reality. As will be argued elsewhere, this and other related facts are certainly not to be considered as happening just by chance, but on the contrary as most significant quantum facts.
\par\medskip    
 To complete this subsection another deep result must be recalled. This result is the strong version of the \textit{free will theorem} \cite{2009}. Based on three indisputable axioms of physics, the fascinating statement of this theorem can be stated as follows. \textit{The answer of a unit spin particle to a measurement of its squared spin along three mutually orthogonal axis, is free.} This means that the measured particle's answer cannot be described by any function of the Universe state, prior to this measurement. 
 \par
 In 2006, a previous version of the theorem had established this freedom in the case of the past light-cone of summit the spacetime point at which the measurement is performed~\cite {2006}. In this case the particle's freedom is with respect to what is called \textit{locality}, \textit{i.e., the einsteinian causality} which is associated to the existence of a universal speed limit in the Universe, concerning energy and information transfers. Quantum correlations though have long displayed non-local aspects. In the relativistic $QFT$ context for example, a simple free field \textit{propagator} (no interactions) exhibits non-local correlations of quantum fields, and the image is sometimes proposed of quantum correlations \emph{leaking out} of the light-cones which delimit locality.
 \par
 This is why the theorem's strong version is so important to definitely set up the absolute freedom a particle has not only with respect to einsteinian causality but also to any possible form of quantum correlations. Whereof the recognition (and `high-tech' exploitation \cite{Gisin}) of a genuine randomness in particle's answers.
 \par\medskip
 There are far reaching consequences of this theorem at the forefront of which, again, that there cannot be any spacetime mechanism that would determine the particle's answers (for example, the collapse in a given measured spin state), and in the same line, that \emph{hidden variables} alternatives are excluded by the same token. To enforce this understanding of things it is worth noting that the same conclusion is reached also from the \emph{quantum information} point of view \cite{Brukner}. Lastly and even more decisive, the weirdness of $QM$ has received a cogent experimental support opposed to the long standing `de Broglie attempts' at reducing $QM$ to some classical description while the latter had been recently reshuffled by the famous Yves Couder's experiments \cite{science}. 
       
 \subsection{\label{D} What is the wave function collapse?}
        
        Out of these considerations, made on the bases of experimental and theoretical physics, it seems reasonable to conclude as follows. 
        \par\medskip
        (i) Like the wave function itself, the collapse is a reality of the quantum system described by the wave function, and not simply a knowledge one would keep updating about the system.
        \par\medskip
        (ii) Though real, the collapse is not to be conceived as a physical mechanism taking place in spacetime, and so, it cannot be accounted for by any non-linear extension to the standard Schr\"odinger equation or other equation. This means that any search of such a mechanism is doomed to failure. 
        \par\medskip
        (iii) It is the more so as the collapse is instantaneous, as seems to be favoured by covariant quantum field analyses. The current understanding of this issue goes as follows. So long as the wave function does not undergo any collapse, it supports (the unitary evolution of) potentialities, which, not being actualized, are localized neither in space nor in time. The collapse comes about with a given actualization of the available possibilities. Along the set of of \emph{worldlines} associated to a quantum set (system plus detecting device), this actualization is tagged in a relativistic invariant way by a corresponding set of \emph{proper-times}. This just defines \emph{instants}, a set of points on the worldlines. Now there is no set of proper-time data that would be antecedent to the collapse (otherwise the wave function would have collapsed already), and thus no invariant collapse duration either. This is why the collapse, transition from a potential state to a state of a higher degree of actualization, can only show up in an instantaneous manner or, at the very least, in a way such that no invariant duration can be ascribed to it. A byproduct of this analysis is that an instantaneity conceived as a very brief duration of time is automatically discarded.
       \par\medskip
       (iv)  In view of the strong free will theorem, and besides serious difficulties with relativistic covariance, alternative mechanisms based on \textit{hidden variables} are therefore to be excluded. The `free will' of `entangled particles' is elementary, irreducible and define a pure randomness~\cite{Gisin}. As things are taken from the side of quantum information theory, moreover, the same conclusion follows again \cite{Brukner}. Eventually, one may also evoke experiments performed with the help of \textit{Hardy-Jordan states} which strongly support $QM$ in its non-local and non-realistic (\emph{i.e.}, potential) aspects, while violating the predictions of local hidden variables theories by 45 standard deviations \cite{Gerry}.
       \par\medskip
       
(v) If the wave function status is as proposed in the present paper (Section IV), then the wave function collapse is, like the wave function itself, more than an updating of the knowledge about the quantum system it is meant to describe. It can only be thought of as \textit{a transition from a state of potency to a state possessing a higher degree of actuality}, and this nails down the collapse proper ontological status. One may again observe that the formalism itself sticks to the situation, as the transition is described within an abstract functional Hilbert space, and not in spacetime where a mechanism could be envisaged. 
       \par\medskip
       (vi) The collapse/transition is caused by some action exerted on the quantum system by a reality being in a higher degree of actuality than the quantum system on which it operates. It is significant to note that no measure performed on quantum systems, quantum non-demolition measurements included, would constitute an exception to this rule \cite{dAriano}.
         \par\medskip
         (vii) Along these lines of thinking the quantum reality, the notion of collapse can be separated from the notion of reduction. This is because, so conceived, the wave function collapse will preferably refer to a genuine physical/ontological change in the quantum system described by the wave function. The wave function reduction instead, may be exclusively relative to some knowledge updating about the system, and thus be relevant to the general case of \emph{a preparation}. As is made explicit in the next Section, in effect, a wave function reduction may be exclusively epistemological.

  \section{\label{SEC:5a} A few revisitations}     
       
 \begin{itemize}      
\item In the fully epistemological Section \ref{SEC:2} a wave function `collapse' without interaction is presented as a rather curious case \cite{Laloe}. Given a particular quantum system, as simple as the quantum optic ones considered in Section \ref{SEC:2}, there is no difficulty to admit that the physicist enumerates the catalog of possibilities the photon can possibly realise \textit{in space} (the various branches it can propagate through), and \textit{in time} (the effective updating of possibilities, once some of them get discarded; if, \textit{e.g.}, $D_1$ registered no click after some long enough duration of time). Now, leaving the detecting enclosure $D_1$, the only `collapse' which has taken place is a collapse in the catalog of possibilities, what is to be properly understood as an updating of the available knowledge on the system. But so long has no click has been registered by $D_2$, no real physical collapse, in the sense proposed here, has ever taken place. And thus, the only reasonable interpretation of this curious experiment is that it is nothing but a preparation/reduction of the system in the `post-detection by $D_1$-state', see FIG.3. Given a quantum entity what is a preparation if not a reduction of all of the possibilities it may display to a few selected ones, allowing for an explicit writing of the wave function?

\begin{figure}[!htb]
 \includegraphics[width=\linewidth]{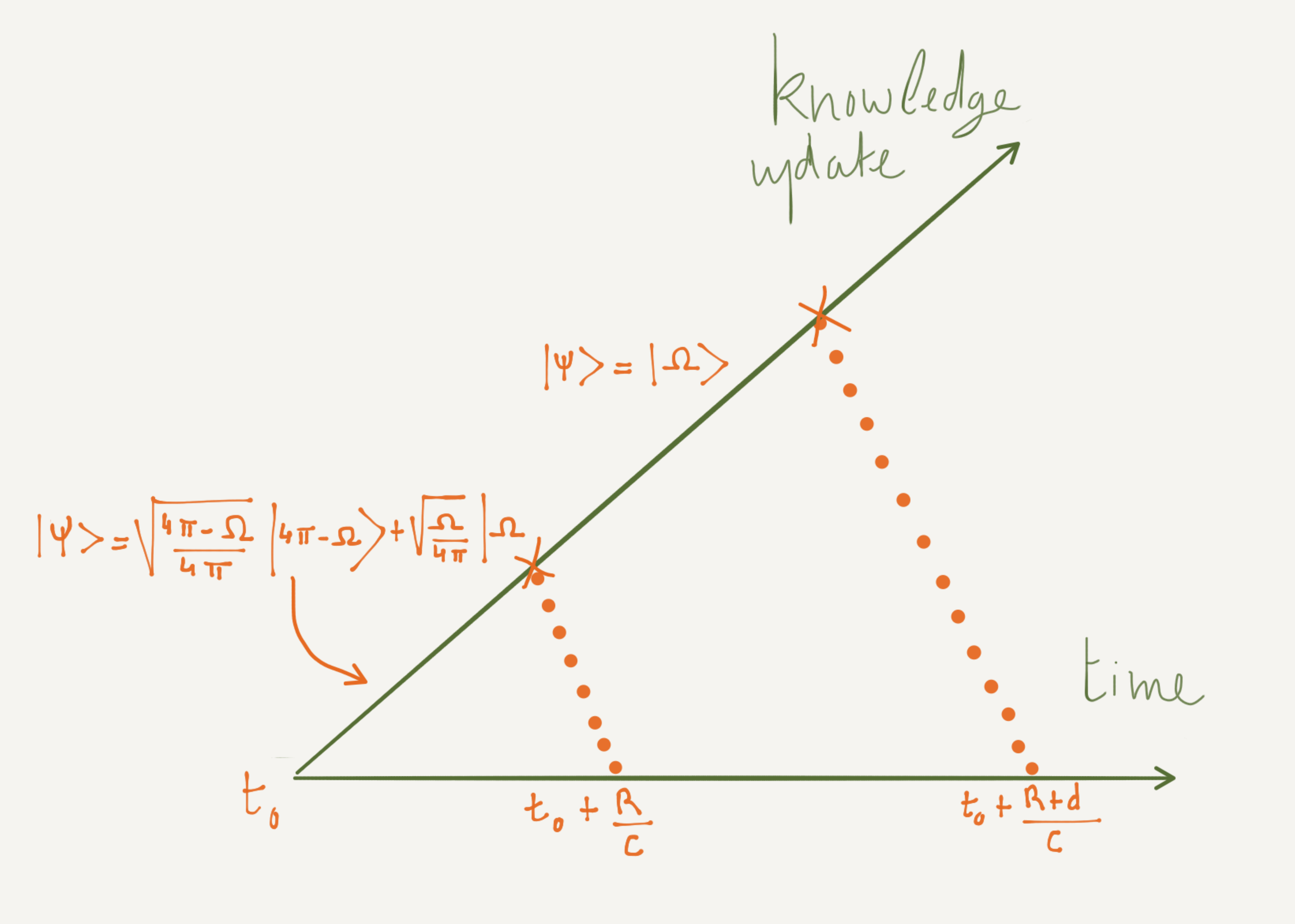}
   \caption{Example of the purely epistemological collapse corresponding to the case pictured in FIG.2 when the emitted radiation has not been detected by $D_1$. In the time interval $[t_o,t_0+R/c]$ the wave function is given by Eq.(5), and beyond, $t\in]t_0 +R/c,t_0 +R+d/c[$, by $|\Omega\rangle$ alone. At the formal level this is a pure collapse. Now it takes place in the knowledge acquired on the system, not within the system itself.}
   \label{epistemocollapse}
\end{figure}

\par
The famous case of \emph{Wigner's friend} can be revisited in a similar way. Inside a closed laboratory, Wigner's friend operates upon a quantum system and can get one of two possible outcomes, either $A=+1$ \emph{or} $A=-1$, say. If the measure gives $A=+1$ for example, then Wigner's friend will describe the system with the help of the state vector $|A=+1\rangle$, while outside of the closed laboratory, Wigner will keep describing the system composed of his friend plus the quantum system, with the help of a superposition of the two possible results, $A=+1$ \emph{and} $A=-1$. It is only by entering the laboratory and by taking notice of the result $A=+1$ that a collapse happens to the Wigner's state vector description of the whole experimental setup. It would seem thereof that the state vector description is relative to the observer and is deprived of any absolute content. As noted in \cite{Laloe}, this has even lead some authors to think that both state vector descriptions should be considered on an equal footing during the intermediate duration of the experiment \cite{Hartle}.
\par
In the proposed understanding of the quantum matter the two descriptions are not taken on the same footing. For Wigner's friend, the first reduction to a state vector described with $|A=+1\rangle$ is indeed a plain physical collapse, endowed with an absolute content.  In the Wigner's state vector description instead, the reduction is nothing but a plain epistemological reduction, a pure updating of Wigner's knowledge, and it is deprived of any absolute character, as can be figured out easily.

\item The case of Section \ref{SEC:3} can be revisited also. There, the ontological dimension of the wave function is claimed to be established because we can measure it through all spacetime, along the lines of Equation (6). Besides the demonstrated impossibility of such a measurement on a single quantum system \cite{dAriano}, this is of course a naive claim. As made obvious by Equation (6), in effect, what is measured is not the wave function $|\Psi\rangle$ itself, but its projection on a given state of localization, as testified again by the formalism itself, \emph{i.e.}, $\Psi(x)=\langle x|\Psi\rangle$, that is a projection of the wave function $|\Psi\rangle$ on one of its possible actualisations. This is but an illustration of the principle posited in the second item of Section IV.B.
 \begin{figure}[!h]
 \includegraphics[scale=0.5]{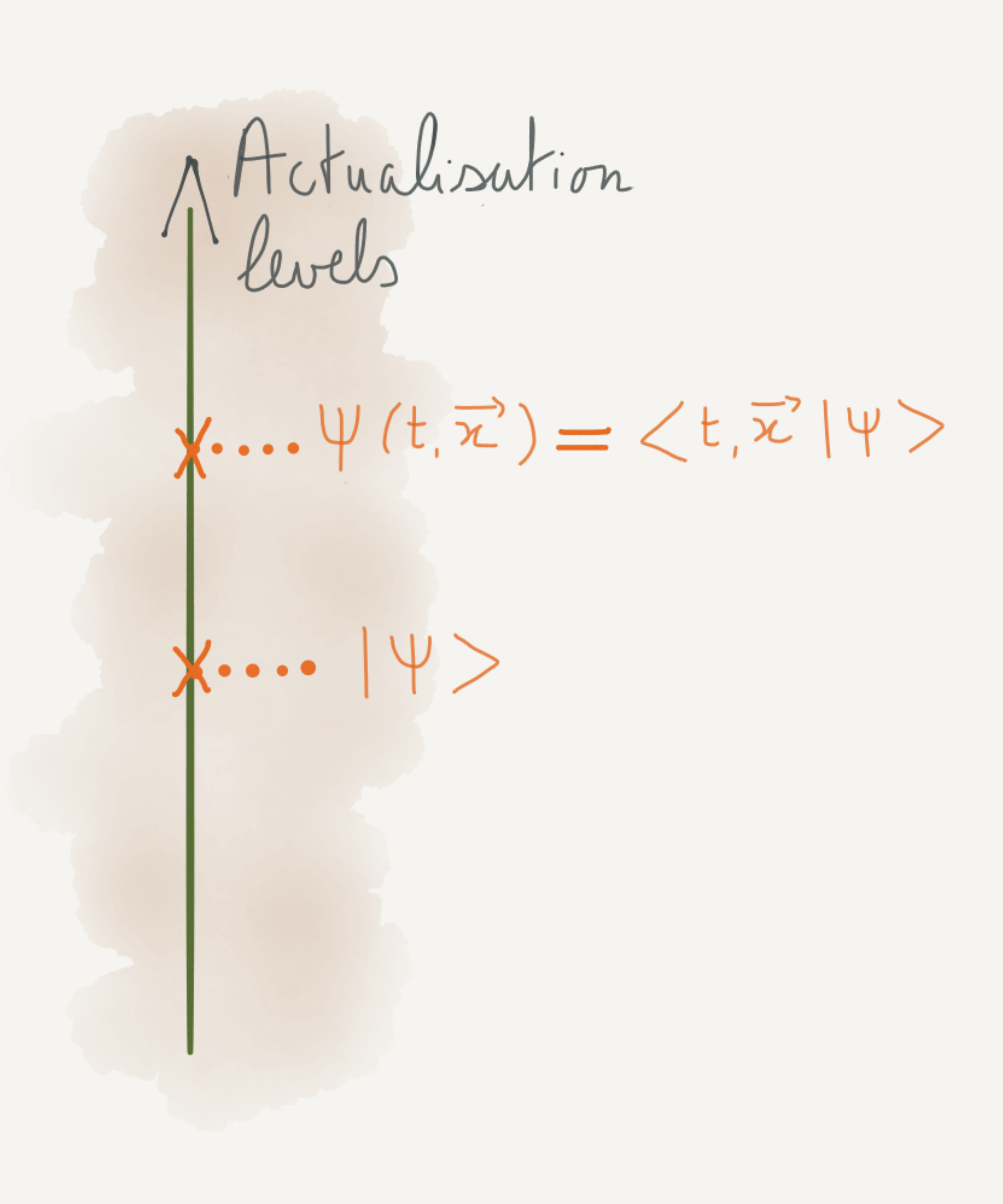}
   \caption{A collapse taking place in the quantum reality itself, here to a given state of localisation in space $|\vec{x}\rangle$ at time $t$, summarised by a state vector $|t,\vec{x}\rangle$ with some abuse of notation ($t$ is not an hermitian operator with associated eigenstates).}
   \label{collapse}
\end{figure}

\item Another agreed on claim can be revisited in the light of the proposed interpretation of the wave function, namely that \textit{..there is no unknown quantum states} \cite{private}. It should be clear enough now that this statement should be restricted to \emph{preparations} only, and one recovers in this way the proper extent of the Introduction second quotation \cite{Grinbaum}.

\item  Some remarkable experimental results are worth being mentioned here because they were heralded as a
progressive collapse. They have been obtained in \emph{cavity quantum electrodynamics} and display the case of a \textit{progressive collapse} of some initial \emph{coherent state} of light into a Fock state (\emph{i.e.}, a state containing a definite number of photons). The converging evolution to the Fock state is caused by a flux of circular Rydberg atoms of \emph{Rubidium}  sequentially sent across the \emph{high Q Fabry-Perot cavity} where the initial state of light is stored (the same strategy is used also in \cite{Bauer} where the converging evolution can be given the mathematical formulation of a stochastic \emph{Markovian} process). This converging evolution to a \emph{steady} Fock state (steady, in comparison to the relaxation time of the cavity) is given the name of a collapse and in this situation it would appear that a duration could be ascribed to the collapse. 
\par
As explained by the authors themselves, however, this progressive collapse to a given Fock state comprising $5$ photons, say, is the cumulative effect of a series of more elementary steps, the \emph{quantum jumps} caused by each atom of Rubidium `measuring' the light state. Once the steady Fock state is reached, the cavity relaxes through quantum jumps. The light state passes from the Fock state of $5$ photons, $|5\rangle$, to the state of $4$ photons and so on, down to the vacuum state $|0\rangle$, at $n=0$ photon. In this way the diagram of the cavity relaxation as a function of time, looks like a perfect staircase. 
\par
However, a zoom taken at the staircase steps would display a quantum jump duration of about $0.01sec.$. Now indeed, these $|n\rangle$ to $|n-1\rangle$ elementary quantum jump durations cannot be attributed to the quantum jumps themselves, but in full rigour, to the duration required by the Rubidium atoms to `realise' that a change has occurred in the light state they are probing by travelling throughout the cavity \cite{Guerlin}.
\par
That is, from the point of view of these very \emph{high resolution} experiments, no real duration can indeed be ascribed to the quantum jumps themselves, and no spacetime mechanism conceived thereof, a feature quantum jumps share with collapses. In the current understanding which is here proposed, however, the two notions are differentiated in a natural manner and this is no doubt an interesting point.
\par
In the cavity, in effect, a light Fock state such as $|n\rangle$ is a state \textit{in a capacity/potentiality} of $n$ photons for all $n\geq 1$ (the case of $n=0$ requiring a separate evaluation \cite{PEs}), and contrarily to the case of collapses, the various Fock states all lie at exactly the same level of the physical reality, as pictured in FIG.6.
\begin{figure}[!htb]
 \includegraphics[width=\linewidth]{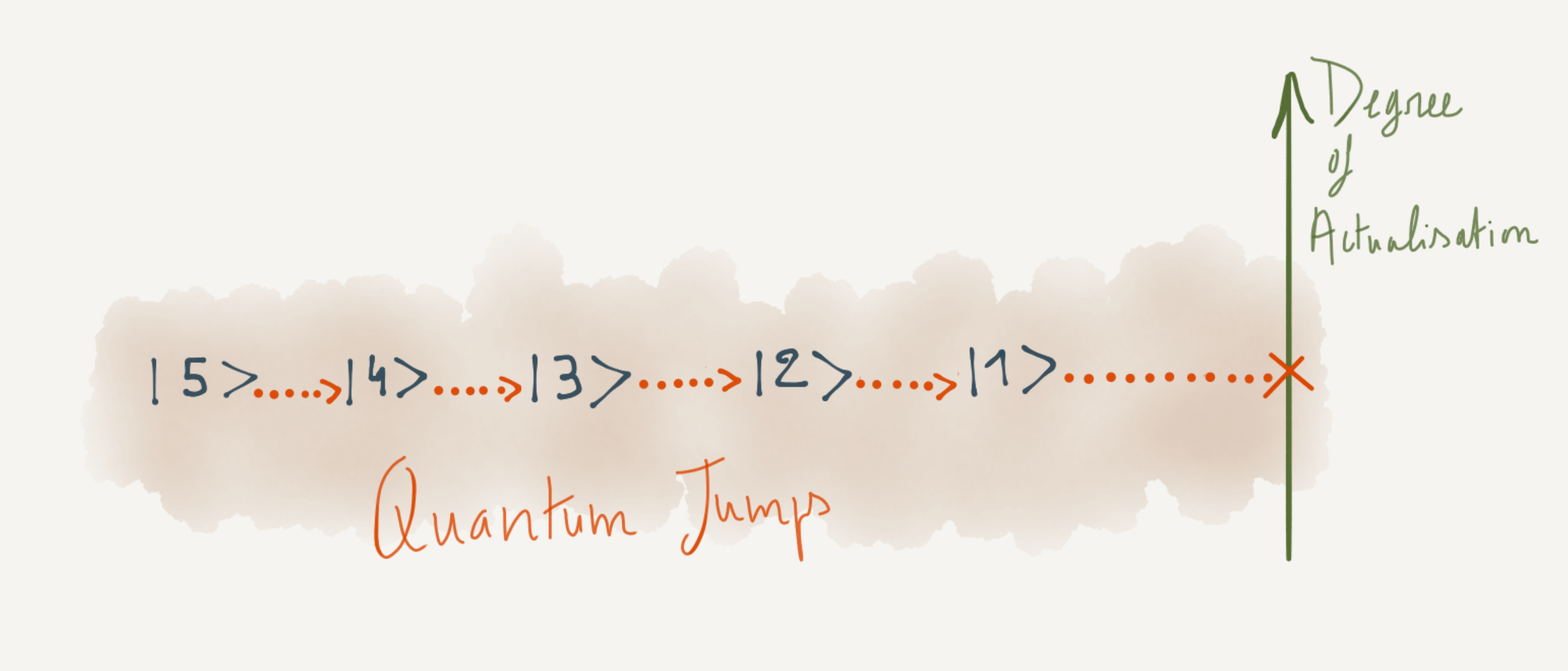}
   \caption{Example of `Quantum Jumps' taking place (within the state vector of the system) at the same degree of actualisation. The Fock state $|0\rangle$ is excluded, as it requires the deeper analysis of Ref.\cite{PEs}.}
   \label{Q-jumps}
\end{figure}

\item Likewise, the quite enigmatic statement \cite{mlb1} that in a \emph{double slit experiment} it is enough to know, just in principle (that is, without it being necessary to perform any concrete detection) which path the photon has followed for the interference fringes to disappear, finds a consistent explanation in the proposed understanding of the matter.

\par
 If, be it in principle only, one is able to know `which path the photon has followed' then, in a way or other the photon wave function has \emph{necessarily} undergone a certain actualisation, directly or indirectly related to its localisation degrees of freedom. Had not it been so, the wave function only described the unitary evolution of potentialities which were localised neither in space nor in time (Section V.D. (iii)), in which case interference fringes did show up; and disappeared otherwise.
 \par
  It is important to realise also that the discourse on the physical reality, \textit{i.e.}, `which path the photon has followed' is but a reconstruction, a reinterpretation \emph{a posteriori} of the experimented physical reality. Before its \emph{click} on the screen located behind the two slits, in effect, it should be clear that no photon did ever propagate through the experimental setup. Note that, within the interpretation currently proposed, this point, widely agreed on nowadays \cite{Gerry}, receives a natural explanation, that is an explanation immediately related to the proposed \emph{nature} of the wave function.

\par Ignoring this last point, the related experiment \cite{Grangier} where photons are sent off to the interferometer sequentially with arbitrarily long lapses of duration, while reproducing the standard pattern of interference fringes, remains poorly interpreted \cite{mlb5}.  This experiment in effect, could suggest that on the average, photons know where to hit the screen so as to build the interference fringes, an interpretation sometimes advocated by some authors who think of elementary particles as endowed with \emph{telepathy}, or that they are controlled by some \emph{Quantum Angels}, the quantum pendants of classical Maxwell's demons \cite{Suarez}.
\par
With arbitrarily long lapses of duration, what is sent off to the
interferometer is, instead, a series of one-photon wave functions comprising
each, in a potential (and constrained) way, all of the points where the photon
can hit the screen. As a large enough number of one-photon wave functions is
sent off to the interferometer, interference fringes show up in the standard
manner, as the whole set of photon local actualisations is here exhausted in a
progressive way (with electrons, for example, fringes appear clearly enough
after some 70.000 runs \cite{Gerry}), while it is much rapidly realised through
the continuous flux of photonic wave functions. It is therefore possible to
look at things the other way round, noticing that it is because experimenters
are able to prepare the same one-photon wave functions always ({\textit{i.e.},
comprising the same set of potentialities) that the same interference patterns
are obtained in the end, with, as a corollary, that the case of sequential
one-photon wave function runs, sent off at arbitrarily long lapses of duration,
has exactly no bearing on the matter. The situation is quite similar to playing
dices, so long as the dices are all the same. That the dices be played at
arbitrary lapses of duration has no influence at all on the resulting
probability distribution of the possible outcomes, when a sufficient number of
runs is achieved. Note in addition that a simple relativistic argument would
deliver the same conclusion.

\item On the so long debated role of the observer in $QM$. In November 2017 an
  article published in the review \emph{New Scientist} had title `Reality? It's
  what you make it' \cite{Ball}. It isn't so. The observer does not
  \emph{decide} of the quantum reality. In agreement with the Introduction
  second quotation, the experimental device an observer sets up constrains the
  realisations a quantum entity is capable of. Still, the observer does not
  determine the quantum system's own answer, that is, the \emph{very}
  realisation which will come out of \emph{a given} experimental run. The
  strong free will theorem \cite{2009} has formalized this irreducible freedom
  in the most cogent and indisputable manner. 

\item Another long exposed aspect of quantum entities is their famous
  \emph{wave/particle duality}. Behaving as waves in some contexts and as
  particles in other contexts, both aspects are thought of at the same level of
  reality, what is rendered by the idea of a \emph{duality}. For the sake of
  analogy, the attractive case of a cylinder is proposed sometimes as an
  example. A cylinder in effect is endowed with the appearance of a circle
  along a given perspective and with that of a rectangle along another
  (orthogonal) perspective, both appearances appropriately taken at the same
  level of reality. Now, since a cylinder is neither a circle nor a rectangle,
  to put things in a blunt way, the conclusion goes as follows. The cylinder
  appears like a circle sometimes and as a rectangle some other times but is
  neither a circle nor a rectangle, it is a \emph{something} on its own, that
  must be taken as is, without it being possible to say anything more about it.
\par
 In the end, is this anything more than a quite convoluted way to state that we
 still don't know what we talk about? For sure, that things may be not knowable
 beyond a certain threshold is a statement which has its validity realm. But
 not here and not~yet.

\par\medskip

In itself, that an entity may display corpuscular and wavy aspects according to
the context is in no way problematic \cite{Bachtoldbis}. What wouldn't make
sense, on the contrary, would be to infer that quantum entities \emph{are} wave
and particle together. This was neatly refuted by Heisenberg arguing that, as a
consequence, the wave and particle aspects were to be taken \emph{only} as mere
analogies \cite{Bachtoldbis}. However, this \emph{proviso} is not satisfying
either because experimental facts exhibit physical behaviours which are
authentically those of waves and particles, even serving the purpose of
defining them as such in some instances.
\par\indent Within the present understanding, now, there is no need for such
sterile convolutions because the ondulatory and corpuscular aspects of a
quantum entity are no longer relevant to the same level of
reality/actualisation in general, such as exemplified by a given quantised
field versus the particle associated to this same field: As explicited in
\cite{PEs} in effect, the corpuscular aspect corresponds to most accurate
protocols and definitions in the respective contexts of experiments and $QFT$s
analysis. Moreover, between the wave and corpuscular aspects, original
experiments carried out in Nice a few years ago have been able to display a
full continuum of other possible actualisations \cite{Tanzilli} showing that
the wave and corpuscular manifestations, as such, are not so mutually exclusive
as we first thought they are. And thus the old idea of a wave/particle duality
does not really stick to the situation (in \cite{Tanzilli} this old idea of a
duality is even recognised an inadequacy).
\item Well before Bell's inequalities violations within $EPR$
  (Einstein-Podolsky-Rosen) type experiments were duly established in 2015
  \cite{2015} , $QM$ had acquired the reputation of being \emph{non-local} (the
  excellent presentation of the matter by \cite{Bachtold} is used here in order
  to summarise the state of affairs).World recognised experts testify to this
  nonlocal interpretation of $QM$,
\par
\textit{There is no way to escape non-locality \cite{Peres2}.}
\par
\textit{What Bells offered us is a proof that there exists a genuine nonlocality in the way Nature really works (...). This nonlocality is, to begin with, a characteristic of $QM$ itself, ..\cite{Albert}}
\par
As noticed in \cite{Bachtold} with relevance, this interpretation is quite
surprising because, as supported by experience, just the opposite is true. $QM$
and its covariant and field most fecund generalisation, the $QFT$s (Quantum
Field Theories) are built-in \emph{local} theories right from the onset. On
theoretical grounds moreover, it is worth noting that attempts at proving that
Bell's inequalities violations should necessarily involve nonlocality have
never been successful \cite{Bachtold}.
\par
In $EPR$ type experiments the idea of nonlocality comes about because of an
apparent immediate spooky action at a distance, conflicting with the
fundamental creed of relativity theory that no action could transfer energy or
information at a speed greater than that of light. This situation has given
rise to the widespread belief that indeed, $QM$ and relativity theories can be
conceived as sharing a `peaceful coexistence' only~\cite{Shimony}. 
\par
The situation is far from being so, since on the contrary, each theory calls
for the other, as could be shown elsewhere while having already benefitted from
a publication \cite{Kerner}. A hint to get out of this puzzling case can be
borrowed to \cite{de Muynck}, and though expressed in the framework of an
\emph{empiricist} interpretation of $QM$, it offers some connection to the
proposed understanding of the quantum world, 
\par
\textit{It is only if $QM$ is supposed to refer to an objective reality that the idea of non-locality comes about, \cite{de Muynck}.}
\par
Within the proposed understanding of things, whenever there is localisation,
the purely potential state vector describing a given quantum system,
$|\Psi\rangle$, has necessarily undergone some
actualisation/projection/collapse in the very sense of $\Psi(t,\vec{x})=\langle
t,\vec{x}|\Psi\rangle$. In this way, it enters space-time and falls under the
constraint that no action at a distance should ever be observed.

Having entered space-time through a projection, or collapse, to a given space-time localised state $|t,\vec{x}\rangle$, any $\Psi(t,\vec{x})$ in $QM$, can be referred to as some objective reality. 
Supposing that the whole set of $QM$ physical phenomena are reducible to the above elementary objective realities \emph{only}, then, given the \emph{immediate} quantum correlations at a distance which are observed in Bell's inequalities violations, there is effectively no way to escape the idea of nonlocality. This is what is understood in the quotation given above \cite{de Muynck}.
\par
Now, locality holds instead. This is asserted within the proper framework where locality considerations are consistently carried out, the covariant quantum theory of fields. In $QFTs$ in effect, locality is an axiom, the famous \emph{Equal Time Commutation Relations} which are textbook material (\textit{e.g.}, with $\mu,\nu=0,1,2,3$ space-time indices),
\begin{equation}\label{field}
[\varphi_\mu(t, \vec{x}), \pi_\nu(t, \vec{y})]=i\delta_{\mu\nu}\,\delta^{(3)}(\vec{x}-\vec{y})\end{equation} where $\pi$ is the momentum canonically conjugated to the $\varphi$-field, and where the other commutators, $[\varphi, \varphi]$ and $[\pi, \pi]$ vanish, \cite{IZ}. The experimentally verified predictions of $QFTs$ attest to the reliability of the relativistic locality axiom. Now, this means that the supposition just made cannot be maintained: $QM$ physical phenomena \textit{are not} generated by the space-time projections $\Psi(t,\vec{x})$ \emph{only}, but also by the `non-objective' realities of fully potential wave functions, $|\Psi\rangle$, as are linear superpositions of orthogonal state vectors, for example. As known since the early days of $QM$, quantum phenomena (such as patterns of interference fringes, to begin with) cannot be described without them.

\par
There is no doubt room for nonlocality in quantum theories and the two opening quotations \cite{Peres2, Albert} make sense. Not full sense however, because the locality and nonlocality aspects are, again, taken at the same level of the total physical reality  with, as a corollary, the unavoidable still fallacious idea of a `peaceful coexistence' only. 
\par
 Within the proposed understanding of the quantum world, on the contrary, the nonlocality aspect of $QM$ has a domain which is identified. It is a sound and real aspect of the quantum domain, whose proper level of reality must be stated as follows. It is the level of reality of $|\Psi\!\rangle$, that is the one of wave functions/state vectors having undergone no space-time actualisation. This level of the physical reality having no projection/actualisation in space-time, doesn't fall under the scope of relativity theory, from where follows the commonly shared impression that `quantum correlations seem to emerge out of an elsewhere of space-time'.
  From where follows also, and most importantly, that there is no contradiction with relativity theory which rules the invariance of physical laws under a group of space-time point transformations of the \emph{affine} Minkowski space \cite{tg}.
 \par
 It is true that the words themselves, locality versus nonlocality, may induce some confusion in that they seem to refer to aspects that could be relevant to one and the same level of the physical reality, as advertised above. 
 \par
 What is to be understood is that nonlocality does not refer, in space-time, to some nonlocal behaviour, what would precisely be suggestive of spooky actions at a distance. Rather, it refers to a realm where there is no localisation \emph{at all}, and this is a radically different situation. Spooky actions at a distance are automatically discarded, as are differentiated the domains where locality and nonlocality apply, without any risk of conflict.

}
\end{itemize}

\par\medskip
 As will be proposed and extended in next papers, a lot of cases can be revisited in the light shed by this interpretation of $QM$, as \textit{mutatis mutandis}, it extends to the deeper level of $QFTs$ \cite{PEs}. In the end, it is the \textit{nature} of the quantum entities which comes progressively to light, and completes the full mathematical description already at disposal. To put things in the words of Section IV$\!.$A, one begins to know in a better way the nature of {`what we talk about'}, and accordingly, to unravel a bit of the quantum enigma.

\section{\label{SEC:5} Elements of a Conclusion}
 `What is Quantum Mechanics trying to tell us?' did D. Mermin ask twenty years ago~\cite{Mermin}. What $QM$ keeps crying out is that what is potential is not deprived of reality, and one could be tempted to say that $QM$ is incredibly obstinate on this point. The matter is that this point totally escapes our common sense so as to make of the superposition principle \textit{the only mystery}, to recall again the famous words of R.P. Feynman.
 Now, why is it so?
  \par\medskip
   The current paper allows one to shed light on this long issue. 

 - For one part, it is clear that superpositions are thought of in a literal manner, as being superpositions of physical actualities. But this cannot make sense. At the formal level of $QM$ which is always found to stick so close to the quantum experienced reality, this is emphasized by the fact that in superpositions, coefficients are necessarily complex-valued \cite{Bachtold}. The origin of the confusion is easily traced back to a simple, still overlooked (metaphysical) point that potency can never be apprehended in itself, but only through the actual realisations it is capable of. That is, potentialities can only be described in terms of actualities, while non actual realities are meant and thereby called potential. This point is fundamental.
 \par
 - For the other part, the origin of the superposition principle mystery is obviously related to the fact that in physics, the reality of potency is simply denied. This is why, along this point of view, \textit{the wave function is the knowledge we have about the system, and nothing else \cite{private}}. One mistakes the wave function and the infinite variety of its potentialities, for the preparations one achieves out of it, which of course are known by construction/selection. As recalled in Section \ref{SEC:3}, in effect, only what is measurable is taken for physical/real, and not so long ago, this attitude was again perfectly coined by D. Mermin's famous quotation that \textit{`correlations only have a physical meaning/reality, what they correlate do not..'} \cite{Mermin}.
 \par\medskip
Not surprisingly then, the combination of the above two attitudes precludes any resolution of `the superposition principle mystery' and preserves the quantum enigma inviolability. 
 \par\medskip 
 
\noindent In the current paper it is claimed that one cannot escape the necessary distinction between a reality and the modes of reality which, to various degrees, can be either actual or potential not to constantly be confronted to the apories and paradoxes of which the literature amply testifies. This distinction for instance becomes essential when it comes to compare classical and quantum physics

 \par
 Positing the proper reality of the potentialities expressed in the wave function (\textit{mutatis mutandis}, in the quantised fields \cite{PEs}) it is claimed that the corresponding mathematical expression refers to a genuine reality of the quantum system it describes, in the same way as the mathematical symbols $\vec{E}$ and $\vec{B}$ refer to genuine electric and magnetic fields in classical physics. Now, whereas the latter enjoy a plain, actual, stable reality on their own, the wave function does not. Its reality, imprinted of potency, is still \emph{incomplete} in a sense which will be made explicit elsewhere \cite{PEs}.
 \par
 This point of view clarifies the epistemological/ontological debate, as the two aspects can no longer be thought of as being antithetic. Nor would one have, \textit{a fortiori}, to discard one of these two comprehensions to the exclusive benefit of the other. Quite on the contrary.  If, as claimed in the current paper, the wave function is the mathematical expression of a genuine physical reality of the quantum system it describes so well, and if this physical reality is that of a certain potency, then, the catalog/epistemological dimension of the wave function follows necessarily, because a potency can only be apprehended through a catalog of actual realisations. 
 \par
 Conversely, if a catalog is that of a given set of potentialities whose probability amplitudes of actualisation, squared, add up to unity under unitary and non-unitary (collapse) evolution, then this catalog of possible actualisations  is given in reference to a conserved \emph{substratum entity}, a \textit{thing}, and this testifies to the ontological dimension of the wave function. To put things in a simple way, no experimentalist would ever think that he or she experiments \textit{no-thing}.

 \par\medskip
 The nature of the much debated collapse of the wave function follows in a natural enough manner with, as an important corollary, that it cannot be described by any sort of spacetime mechanism, and as a by-product, that the three notions of collapse, reduction and quantum jumps can be differentiated in a relevant way. These considerations are developed on the basis of a few principles, only four of them, which allow one to unravel a certain number of long standing quantum issues as quickly sketched in Section VI, in a short list of eight examples. In \cite{PEs} these principles will be extended so as to cover the whole spectrum of elementary particles we know from the experimentally established standard model of elementary particles. 

\begin{acknowledgments}
It is a pleasure to thank Br. G.-M. Grange, O.P., for a careful reading of the manuscript 
\end{acknowledgments}


\begin{thebibliography}{}
\bibitem{Laloe2}
An instructive account of these experiments is given in \cite{Laloe}, Chapter 7, in particular the Reference [307] therein.
\end{thebibliography}
\end{document}